\documentclass[sigconf,10pt]{acmart}
\AtBeginDocument{%
 }

\renewcommand\footnotetextcopyrightpermission[1]{}

\usepackage{algorithmic}
\usepackage{graphicx}
\usepackage{textcomp}
\usepackage{xspace}
\usepackage{url}
\usepackage{booktabs}
\usepackage{tabularx}
\usepackage{pifont}
\usepackage{listings}
\usepackage{subcaption}

\definecolor{purecowrie}{HTML}{D9D9D9}
\definecolor{cowriehp}{HTML}{A6A6A6}
\definecolor{pureondevice}{HTML}{C9E0E8}
\definecolor{ondevicehp}{HTML}{6FA8BE}
\definecolor{purecloud}{HTML}{B7C4DB}
\definecolor{cloudhp}{HTML}{4C6FA0}
\definecolor{hydraorange}{HTML}{E07B39}

\newcommand{\legswatch}[1]{\raisebox{-0.05em}{\colorbox{#1}{\rule{0pt}{0.75em}\rule{0.75em}{0pt}}}}

\newcommand{\acro}{\textsc{HydraPoT}\xspace}

\begin{document}

\title{HydraPoT: A Configurable Multi-Agent Framework for Cost-Aware LLM-Assisted Honeypots}

\author{Kamolchanok Saengtong}
\affiliation{%
  \institution{Prince of Songkla University}
  \city{Phuket}
  \country{Thailand}
}
\email{s6630613011@phuket.psu.ac.th}

\author{Chutirat Saengyingyongwattana}
\affiliation{%
  \institution{Tsinghua University}
  \city{Shenzhen}
  \country{China}
}
\email{hpp26@mails.tsinghua.edu.cn}

\author{Sasin Janpuangtong}
\affiliation{%
  \institution{Chiang Mai University}
  \city{Chiang Mai}
  \country{Thailand}
}
\email{sasin.ja@eng.cmu.ac.th}

\author{Norrathep Rattanavipanon}
\affiliation{%
  \institution{Prince of Songkla University}
  \city{Phuket}
  \country{Thailand}
}
\email{norrathep.r@psu.ac.th}

\renewcommand{\shortauthors}{Saengtong et al.}
\begin{abstract}
Large language models (LLMs) can improve honeypot realism by generating dynamic, context-aware responses, but
powerful cloud models incur substantial cost and latency,
while lightweight models may provide weaker deception.
This forces practitioners to choose between operational efficiency and interaction fidelity.
In this work, we present HydraPoT, a configurable multi--agent honeypot framework that integrates traditional emulation, lightweight LLMs, and powerful cloud LLMs. HydraPoT routes each attacker command to an appropriate
agent according to a configurable complexity-based policy, while maintaining shared session state to preserve consistency across agent transitions. We evaluate HydraPoT through an architectural-overhead analysis, a cross-agent
context-preservation benchmark, and a three-agent case
study. HydraPoT introduces less than 5\% runtime overhead,
preserves context across heterogeneous agents, and achieves
fidelity close to a cloud-LLM-only honeypot. According to our case study policy, it reduces cloud-processed tokens by 72.6\%
and operational cost by 3.7×. These results demonstrate that
configurable multi-agent routing can make realistic LLMassisted honeypots more practical under limited operational
budgets.
\end{abstract}


\maketitle


\section{Introduction}
Cyberattacks continue to grow in scale and sophistication, with ransomware, malware, social engineering, denial-of-service, and supply-chain attacks remaining major threats~\cite{enisa2025threatlandscape}. Their impact extends beyond digital services to physical infrastructure, as demonstrated by Internet-scale denial-of-service attacks~\cite{cloudflare2025ddos} and attacks on industrial and cyber-physical systems~\cite{karnouskos2011stuxnet,abrams2008malicious}. Defending against these threats requires practical insight into how attackers discover targets, exploit services, and interact with compromised systems, making threat intelligence an important component of modern cybersecurity~\cite{valeros2023attacker,franco2021survey}.

Honeypots support this objective by exposing decoy services in controlled environments and recording attacker interactions without placing production systems at risk. They are commonly classified as low/medium/high-interaction systems~\cite{franco2021survey}. Low-interaction honeypots (e.g., Honeyd~\cite{provos2003honeyd}) are lightweight but provide only limited predefined responses. High-interaction honeypots expose functional systems and offer greater realism, but are more difficult and risky to operate. 
Medium-interaction honeypots balance these trade-offs by emulating richer services while restricting attacker capabilities, e.g., Cowrie~\cite{oosterhof_cowrie} emulates SSH/Telnet services and records attacker activity within a simulated shell.

However, traditional medium-interaction honeypots rely on predefined emulation logic and may produce unrealistic responses to unexpected, context-dependent, or complex commands. Recent work addresses this limitation using large language models (LLMs) to generate dynamic and context-aware shell responses~\cite{sladic2024llm,guan2024honeyllm} and to manage long interaction histories through prompt and memory management~\cite{wang2024honeygpt}.

Nevertheless, existing LLM-assisted honeypots rely on a single LLM configuration or a fixed processing strategy. 
\emph{Consequently, security practitioners must often choose between powerful cloud-based LLMs, which provide higher deception fidelity but incur greater cost and latency, and lightweight models, which reduce cost at the expense of response quality.}

To address this limitation, we propose \acro, a configurable multi-agent honeypot framework that integrates traditional honeypot emulation, lightweight LLMs, and powerful cloud-based LLMs within one environment. 
At its core, \acro provides a configurable routing mechanism that selects an agent according to the complexity of each attacker command and a user-defined routing policy. 
This allows practitioners to configure the trade-off between operational cost and deception fidelity; for example, simple commands can be routed to lower-cost agents while powerful cloud LLMs are reserved for more complex interactions. Since consecutive commands may be handled by different agents, \acro also maintains shared session state to preserve consistency across agent transitions.

We evaluate \acro through an architectural evaluation and a three-agent case study. Our results show that \acro introduces less than 5\% runtime overhead w.r.t. the corresponding standalone LLM agents and preserves session context across heterogeneous agent transitions. Under a case-study routing policy using Cowrie, an local LLM, and a cloud LLM, \acro achieves higher response fidelity than the Cowrie-only and local-LLM-only baselines while substantially reducing operational cost compared with the cloud-LLM-only baseline. These results demonstrate that configurable multi-agent routing provides a practical cost--fidelity trade-off for LLM-assisted honeypots. 
We open-source \acro implementation,  which is available on GitHub~\cite{saengtong2026hydrapot}.

In sum, we make the following contributions:

\begin{itemize}
    \item We propose \acro, a configurable multi-agent honeypot framework that integrates traditional honeypot emulation, lightweight LLMs, and powerful LLMs within a single controlled environment.

    \item We design a configurable, state-preserving routing mechanism that selects an appropriate response agent according to command complexity and a user-defined policy, enabling practitioners to control the trade-off between operational cost and deception fidelity.

    \item We evaluate \acro through architectural-overhead, cross-agent context-preservation, and cost--fidelity experiments, showing that it introduces low overhead, maintains session consistency across agents, and reduces operational cost while retaining fidelity close to a cloud-LLM-only honeypot.
\end{itemize}

\section{Related Work}

LLM-based honeypots have emerged since 2022~\cite{bridges2025sok}. Early studies demonstrated that general-purpose LLMs could emulate Linux, macOS, Windows, and common command-line tools~\cite{mckee2023chatbots}. ShelLM~\cite{sladic2024llm}, among the earliest complete implementations, used GPT-3.5 to emulate SSH/Telnet services and showed through a user study that most users could not reliably identify the honeypot. Its extension, VeILMes~\cite{11129519}, fine-tuned Llama~3 8B on Cowrie command traces and expanded support to HTTP and MySQL services.
Subsequent systems improved interaction fidelity/functionality/security by automating MITRE ATT\&CK mapping~\cite{gizzarelli2024honeypot}, improving shell's responses to be more authentic~\cite{10.1007/978-981-97-8801-9_13}, and mitigating against prompt injection attacks~\cite{10679411}. 
Other studies investigated locally deployed open-weight models~\cite{10808265}, compared RAG with prompt tuning for lightweight shell emulation~\cite{11391242}, explored supervised fine-tuning on attacker interactions~\cite{10735607}, and developed automated fidelity metrics for LLM-based SSH honeypots~\cite{10703029}. Although these systems improve realism, invoking an LLM for every command can incur substantial latency and operational cost.

Existing approaches aim to reduce this cost either by limiting which commands reach an LLM or by reducing the context sent with each invocation. 
The work in~\cite{10295397} combined GPT-based terminal simulation with prompt and output management, reducing token usage by up to 77\%. 
HoneyLLM~\cite{guan2024honeyllm} filters scanning traffic before invoking the LLM, whereas LLMHoney~\cite{malhotra2025llmhoney} uses a dictionary-based virtual filesystem to emulate common commands and forwards unseen commands to an LLM. HoneyGPT~\cite{wang2024honeygpt} instead employs prompt management and memory pruning to retain only impactful session history to be forwarded to an LLM agent. 
The work in~\cite{ai7080287} uses deterministic handlers for most commands to confine hallucination, with reduced LLM usage as an additional benefit. 
A separate line of work uses multiple LLM agents to improve response consistency rather than reduce cost. AdvancedShelLM~\cite{sladic2026advancedshellm}, for example, uses a Worker agent to generate responses and a Manager agent to verify them, but still requires an additional LLM call (or even multiple LLM calls) for every command.

In contrast to prior systems that employ a fixed filtering policy, a single LLM configuration, or a fixed multi-agent workflow, \acro provides configurable per-command routing among a traditional honeypot, a lightweight LLM, and a powerful LLM. This allows security practitioners to explicitly control the trade-off between operational cost and deception fidelity. \acro also maintains shared state across agent transitions and supports pluggable components and SIEM integration, enabling it to operate as part of a broader cybersecurity infrastructure.

\section{HydraPoT Design}


\begin{figure*}[!tp]
    \centering
    \includegraphics[width=.8\linewidth]{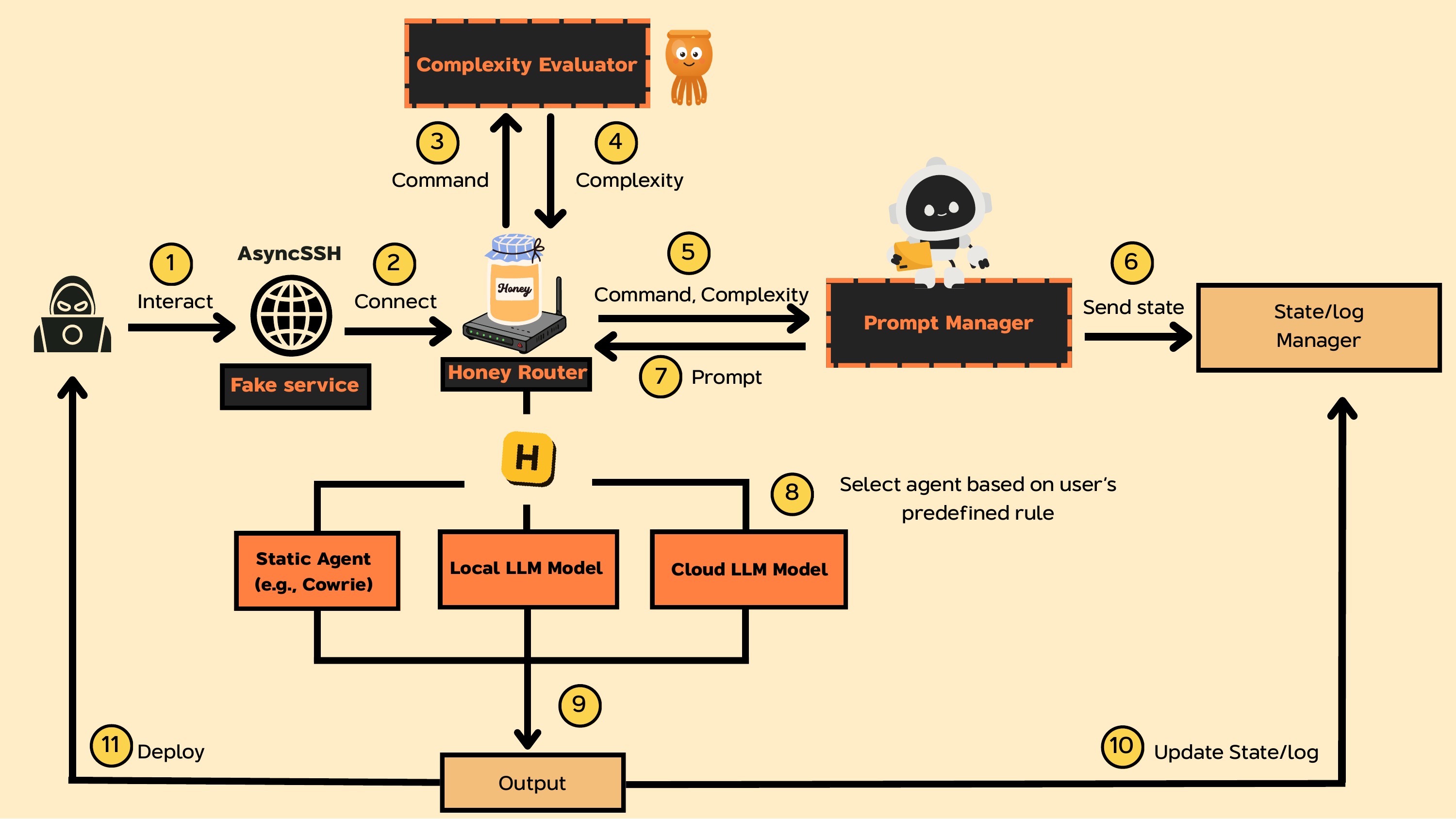}
    \caption{\acro design}
    \label{fig:diagram}
\end{figure*}
\subsection{Design Goals}\label{subsec:goals}
We consider the following goals in \acro design.

\noindent\textbf{G1 --} \textbf{Support for configurable heterogeneous agents.}
    \acro should support different types of honeypot agents within a single framework.
    These agents can broadly be classified into three groups: 
    (1) traditional rule-based (static) honeypots (e.g., Cowrie);
    (2) lightweight/low-cost LLM agents (i.e., local LLMs or inexpensive cloud-hosted models); and
    (3) powerful LLM agents (i.e., large/frontier models hosted in a datacenter).

\noindent\textbf{G2 --} \textbf{Support for configurable routing policies.}
    \acro should allow security practitioners to configure how attacker commands are routed among the available agents.
    The routing policy should provide explicit control over the trade-off between two operational cost and deception fidelity.
    This allows organizations to select a routing configuration that matches their deployment requirements and available budget.

\noindent\textbf{G3 --} \textbf{Cross-agent session consistency.}
    All heterogeneous agents in \acro should maintain a consistent view of the attacker session.
    When the selected agent changes in the middle of the attacker session, relevant honeypot states 
    should remain synchronized across agents.
    This prevents agent switching from producing conflicting responses or inconsistent session behavior.

\noindent\textbf{G4 --} \textbf{Low architectural overhead.}
    The routing, agent-switching, and state-synchronization mechanisms introduced by \acro should incur low additional latency and computational cost.
    In particular, their overhead should be small relative to the cost of invoking the selected LLM agent, so that the framework does not negate the efficiency benefits of using heterogeneous agents.

\noindent\textbf{G5 --} \textbf{Support for threat-intelligence collection.}
    \acro should record attacker commands, generated responses and selected agents.
    These records then can be used for subsequent threat-intelligence analysis, attacker-behavior profiling, and integration with existing security-monitoring systems (e.g., SIEM).

\subsection{Design Overview}\label{subsec:design}

In this subsection, we present a high-level overview of \acro, as illustrated in Figure~\ref{fig:diagram}.
We then describe its critical components in more detail in Section~\ref{sec:detail}.

\acro provides a modular design consisting of five main components: the \emph{Fake Service}, \emph{Honey Router}, \emph{Complexity Evaluator}, \emph{Prompt Manager}, and \emph{State/Log Manager}.
\acro can route attacker commands to three broad types of honeypot agents: a static honeypot agent, a lightweight LLM agent, and a more powerful LLM agent.

Similar to a traditional honeypot, an attacker first establishes a connection with \acro through the front-end Fake Service (\ding{182}), which simply records the received command and forwards it to the Honey Router (\ding{183}).
The Honey Router is the core orchestration component of \acro.
Its responsibilities are to determine which honeypot agent should process the command and to obtain an appropriate prompt for that agent when an LLM-based agent is selected.

To select a honeypot agent, the Honey Router first forwards the attacker command to the Complexity Evaluator (\ding{184}) with the goal of  determining the complexity level of the command.
In this work, we define command complexity based on the impact to the system states, as later described in Section~\ref{subsec:complexity}.
The resulting complexity level is then returned to the Honey Router (\ding{185}).

Based on this complexity level and a user-defined routing policy, the Honey Router selects the honeypot agent that should process the command.
Security practitioners can configure this policy by assigning different agents to different complexity levels.
For example, an organization with a limited budget may route more commands to the static or lightweight agents, whereas an organization prioritizing deception fidelity may route a larger fraction of commands to the higher performing LLM agent.

If an LLM-based agent is selected, the Honey Router forwards the attacker command to the Prompt Manager (\ding{186}).
The Prompt Manager constructs a context-aware prompt that includes the information required to generate a response consistent with the preceding commands in the same session.
To obtain this context, it queries the State/Log Manager for relevant session state across all agents processed the previous commands (\ding{187}).
The Prompt Manager then constructs the final prompt and returns it to the Honey Router (\ding{188}).

At this point, the Honey Router has both the selected agent and the corresponding input (i.e., the original command for the static agent or the constructed prompt for an LLM-based agent).
It forwards the input to the selected agent (\ding{189}), which processes it and generates a response (\ding{190}).

Before the response is returned to the attacker, \acro forwards the command, generated response, and selected-agent information to the State/Log Manager (\ding{191}).
The State/Log Manager determines whether the interaction introduces state that may be required by subsequent commands and updates the shared session state accordingly.
In addition, it records the attacker command, generated response and     selected agent for subsequent threat-intelligence analysis or integration with external security-monitoring systems.
Finally, the Fake Service displays the response to the attacker.

\subsection{Achieving Design Goals}
We now explain how the design presented in Section~\ref{subsec:design} achieves the five goals defined in Section~\ref{subsec:goals}.

\acro supports three broad types of honeypot agents: a static honeypot agent, a lightweight LLM agent, and a powerful LLM agent. These heterogeneous agents are integrated within a common framework and can be selected through the Honey Router, thereby satisfying \textbf{G1}.

To achieve \textbf{G2}, \acro relies on the Complexity Evaluator to estimate the difficulty for a honeypot agent to generate a realistic response to each attacker command. 
Security practitioners can then configure the routing policy within the Honey Router by assigning different complexity levels to different agents. 

To satisfy \textbf{G3}, the State/Log Manager maintains a shared view of the attacker session across all agents. After each command is processed, the State/Log Manager records the interaction and extracts any relevant state changes. When a subsequent command is routed to an LLM-based agent, the Prompt Manager retrieves the relevant shared state 
to construct a context-aware prompt. As a result, different agents can observe the effects of commands previously handled by other agents and maintain a consistent view of the session.

\acro introduces additional processing overhead for command-complexity evaluation, shared-state maintenance, and context-aware prompt construction. However, these components are designed to remain lightweight relative to the cost of executing the honeypot agent. 
In Section~\ref{subsec:overhead}, we show that the architectural overhead introduced by \acro is less than 5\%  under our evaluated configuration, thereby satisfying \textbf{G4}.

Finally, the State/Log Manager records attacker commands, generated responses, selected agents, and relevant execution metadata in a consistent format. These records can be used for subsequent attacker-behavior and threat-intelligence analysis, satisfying \textbf{G5}. 

\section{Details}\label{sec:detail}

In this section, we describe three non-trivial aspects of \acro's design: 
(1) how the Complexity Evaluator determines the complexity level of an attacker command (Section~\ref{subsec:complexity}), 
(2) how the State/Log Manager maintains shared-state consistency across heterogeneous agents (Section~\ref{subsec:shared-state}), and
(3) how Prompt Manager can leverage shared states to construct context-aware LLM prompts (Section~\ref{subsec:context-aware-prompt}).

\subsection{Complexity Levels}\label{subsec:complexity}

We propose to estimate the complexity of emulating an attacker command based on the degree to which the command changes the state of the underlying service.
Our rationale is that stateless commands (e.g., reading/displaying information) are generally easier for a honeypot to emulate, whereas commands that introduce more substantial changes to the underlying state are typically more difficult to emulate and maintain realistically and consistently.

\begin{table}[t]
	\centering
	\scriptsize
	\setlength{\tabcolsep}{3pt}
	\renewcommand{\arraystretch}{1.1}
	\caption{Command complexity based on state impact}
	\label{tab:fi_categories}
	\begin{tabular}{@{}c
			>{\raggedright\arraybackslash}p{0.40\columnwidth}
			>{\raggedright\arraybackslash}p{\dimexpr0.60\columnwidth-4\tabcolsep-1.8em\relax}@{}}
		\toprule
		\textbf{Level} & \textbf{Description} & \textbf{Examples} \\
		\midrule
		$F_0$ &
		Read file, display system information &
		\texttt{ls}, \texttt{whoami}, \texttt{cat}, \texttt{uname} \\
		
		$F_1$ &
		Create file, install tool &
		\texttt{touch}, \texttt{mkdir}, \texttt{wget}, \texttt{curl}, \texttt{apt install} \\
		
		$F_2$ &
		Modify files/dir, change working directory, change shell &
		\texttt{chmod}, \texttt{rm}, \texttt{cd /tmp}, \texttt{mv}, \texttt{sed} \\
		
		$F_3$ &
		Start/stop service, remote access, execute code, elevate privilege &
		\texttt{sudo}, \texttt{systemctl}, \texttt{nmap}, \texttt{curl ... | sh} \\
		
		$F_4$ &
		Impact services, delete files, password changed &
		\texttt{passwd}, \texttt{useradd}, \texttt{chmod 777} \\
		\bottomrule
	\end{tabular}
\end{table}

Accordingly, we use the command \emph{impact factor} ($F_i$) introduced by prior work~\cite{wang2024honeygpt} as a means to approximate the complexity levels.
The impact factor classifies shell commands into five levels, $F_0$-$F_4$, according to their impact on the underlying system state.
At the lowest level, $F_0$ contains commands that read/display information while
the highest level, $F_4$, corresponds to commands that cause substantial or security-critical state changes.
The Complexity Evaluator assigns each incoming command to one of these levels, which is subsequently used by the Honey Router when selecting an appropriate agent.
Table~\ref{tab:fi_categories} summarizes the five complexity levels and provides representative command examples.

\subsection{Cross-Agent State Consistency}\label{subsec:shared-state}

To maintain state consistency across heterogeneous agents, the State/Log Manager performs bookkeeping of state changes within each attacker session. 
Focusing on SSH/Telnet honeypots, \acro tracks six categories of session state: 
(i) the current working directory, 
(ii) files together with their metadata and contents, 
(iii) installed software packages and their versions, 
(iv) user and group information, 
(v) authentication and security settings, such as passwords and permissions, and 
(vi) services and their status.

The State/Log Manager maintains these states in a per-session \texttt{SYSTEM\_STATE} dictionary, which serves as the authoritative shared state for all agents. 
When a new session starts, \acro initializes this dictionary. 
After a command is processed, it receives the attacker command, its assigned impact factor $F_i$, and the generated response from Step~\ding{191} in Figure~\ref{fig:diagram}. 
It then identifies which state categories are affected and updates the corresponding entries in \texttt{SYSTEM\_STATE}. 
%
For example, an $F_0$ command results in no state update; whereas, an $F_1$ command may install a software package. In this case, the State/Log Manager extracts the resulting package information from the command-response pair and updates the corresponding entries, i.e., (iii) in \texttt{SYSTEM\_STATE}.

\subsection{Context-Aware Prompts}
\label{subsec:context-aware-prompt}

When an attacker command is received and the Honey Router selects an LLM-based agent, the Prompt Manager retrieves the latest session state from the State/Log Manager through \texttt{SYSTEM\_STATE}.
It then constructs a context-aware prompt by combining this state with the system prompt and the attacker command.
The resulting prompt is sent to the selected LLM for inference, and the generated response is subsequently used to update \texttt{SYSTEM\_STATE}.
This process is repeated for every command throughout the attack session.
As a result, each new prompt reflects the latest system state produced by commands previously handled by any agent, allowing different LLM agents to maintain a consistent view of the session.




\section{Evaluation}

\subsection{Implementation and Evaluation Setup}

We implement \acro in Python, where the Fake Service is implemented using its AsyncSSH library. For the static honeypot agent, we use Cowrie deployed from its official Docker image. 
The powerful cloud-based LLM agent is accessed through an OpenAI-compatible Python client.
For the lightweight LLM agent, we run an local model on our machine in which \acro supports both the \texttt{llama-cpp-python 0.3.34} library \cite{llamacpp_python} for models in the GGUF format and Hugging Face's \texttt{transformers 5.14.1} library \cite{huggingface_transformers_pypi} for other model formats. 
This modular design allows organizations to replace the local or cloud-based LLM agents without modifying the core \acro framework.

We conduct our evaluation on a desktop equipped with an Intel Core i7-11700K processor, 32~GB of RAM, and an NVIDIA GeForce RTX~4060 GPU. For the lightweight LLM agent, we use \texttt{unsloth/Qwen3.5-9B-GGUF} with the \texttt{Q5\_K\_M } quantization format \cite{unsloth_qwen3.5_gguf} with the following hyperparameters: the context window of 8192 tokens, the max generation length of 256 tokens, a temperature value of 0, and a repetition penalty value of 1.15. 
For the cloud LLM agent, we use DeepSeek-V4 Flash \cite{deepseek_pricing} through its official cloud API.

\subsection{Architecture Overhead}\label{subsec:overhead}

\begin{figure}[h]
    \centering
    \includegraphics[width=.8\linewidth]{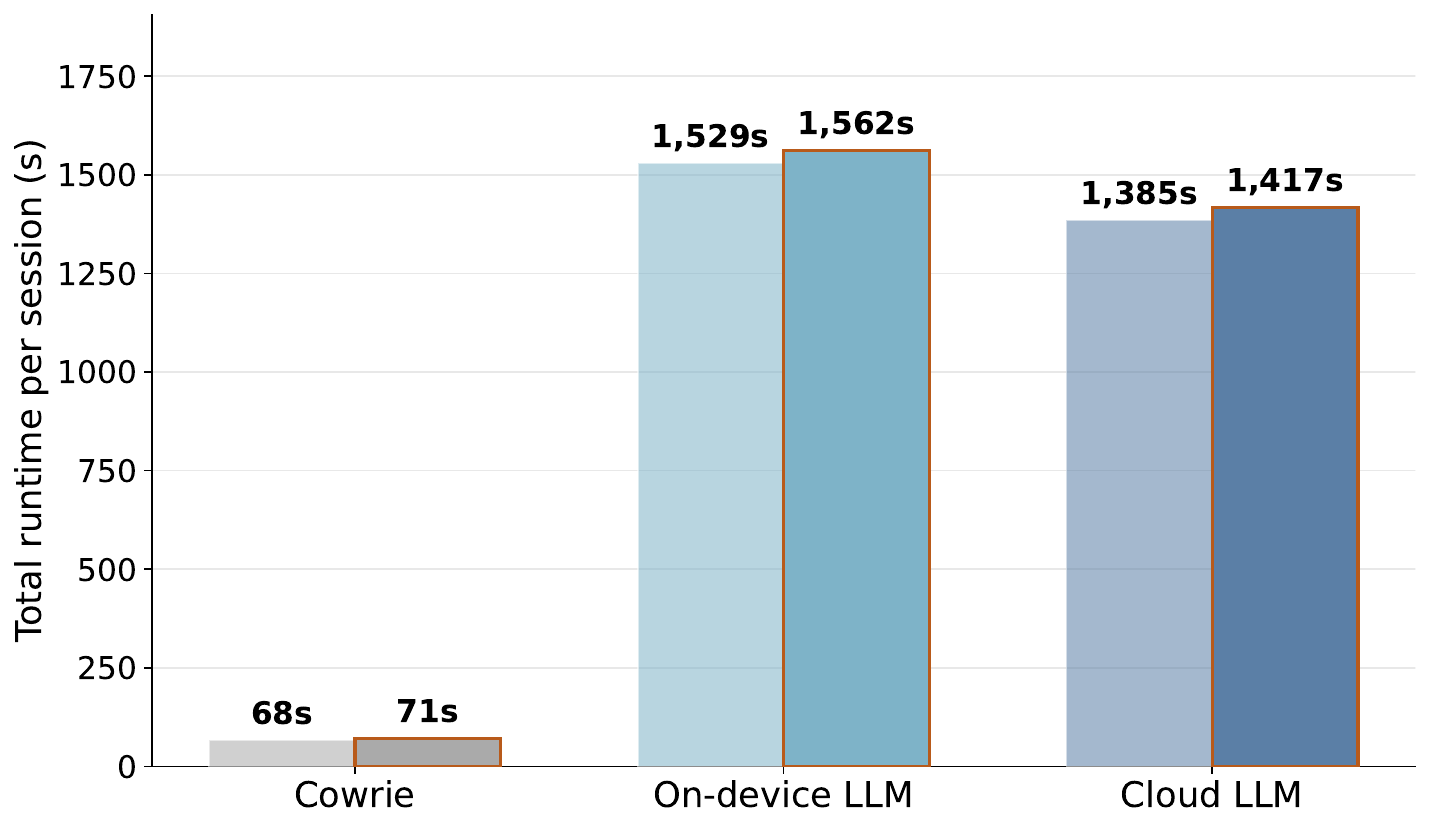}
    \caption{Total runtime of fixed-routing \acro configurations (bars enclosed by orange box) and the standalone agents (bars without orange box); $SD < 1\%$.} 
    \label{fig:avg_routing}
\end{figure}
Here, we evaluate the framework-specific overhead introduced by \acro using three fixed-routing configurations. 
In each configuration, we force \acro's Honey Router to route every command to one pre-defined agent: the static agent, the lightweight LLM, or the powerful LLM. 
All other \acro components, including the Complexity Evaluator, Prompt Manager, and State/Log Manager, remain enabled. 
We compare each configuration with a corresponding single-agent baseline that uses the same agent directly, without \acro's routing and cross-agent state-synchronization mechanisms. 
This setup isolates the overhead introduced by the \acro framework while avoiding variations caused by switching among different agents.

For this experiment, we report the total runtime using the \textit{mrheinen/linux-commands} dataset~\cite{mrheinen_linux_commands}, which contains 480 real-world Linux command records. As shown in Figure~\ref{fig:avg_routing}, \acro introduces only small runtime overhead w.r.t. the corresponding baselines: around 5\% for the static agent and 2\% for the lightweight and powerful LLM agents. 
Moreover, enabling \acro's routing and state-management mechanisms does not increase the peak RAM or VRAM consumption in any of the evaluated configurations.












% ...

\subsection{Context Preservation}

Next, we evaluate \acro's ability to preserve system state across heterogeneous agents using a benchmark of stateful SSH command sequences. Each sequence contains a \emph{context-creating command} that modifies the system state, followed by a \emph{context-dependent command} whose correct response depends on that modification. We construct 29 such command pairs, comprising 58 commands in total from the ML4Net SSH-Shell-Attacks dataset~\cite{botticella2025sshshellattacks}
For comparison, an attacker session in the same dataset contains around 20 commands on average; thus, our benchmark reflects a realistic length of stateful interaction in a typical attacker session.

For each command pair, we force \acro to route the context-creating command to a designated \emph{Creator Agent} and the context-dependent command to a \emph{Reader Agent}.
We evaluate all Creator--Reader combinations except those where Cowrie serves as the Reader, as the Cowrie version used in our experiments does not support updating its internal state based on externally maintained session state, i.e., \texttt{SYSTEM\_STATE}.


{\color{red}\begin{table}[!t]
\centering
\scriptsize
\caption{Context-preservation rate of \acro}
\label{tab:context_preservation}
\begin{tabular}{llccc}
\toprule
\textbf{Creator} & \textbf{Reader} & \textbf{State sync disabled} & \textbf{State sync enabled} & \textbf{$\Delta$} \\
\midrule
Cowrie     & Cloud     & 17.2\% & 100\% & +82.8 \\
Cowrie     & On-device & 79.3\% & 100\% & +20.7 \\
Cloud      & On-device & 14.9\% & 100\% & +85.1 \\
On-device  & Cloud     & 14.9\% & 100\% & +85.1 \\
\bottomrule
\end{tabular}
\end{table}}

We consider each command pair to succeed when the Reader Agent's response matches the ground-truth output obtained by executing the same sequence on an Ubuntu 22.04 VM. Table~\ref{tab:context_preservation} reports the success rates with and without the cross-agent state synchronization mechanism provided by the State/Log Manager, averaged over three runs.
Enabling state synchronization increases the success rate by up to 85.1 percentage points across the Creator--Reader combinations. With state synchronization enabled, \acro achieves a 100\% success rate across all 29 command pairs.
These results show that the shared \texttt{SYSTEM\_STATE} consistently preserves attacker context across agent transitions. 

\subsection{Case study for configurable routing}\label{subsec:case-study}

\begin{figure*}[t]
\centering
\begin{subfigure}[b]{0.3\textwidth}
    \centering
    \includegraphics[width=\textwidth]{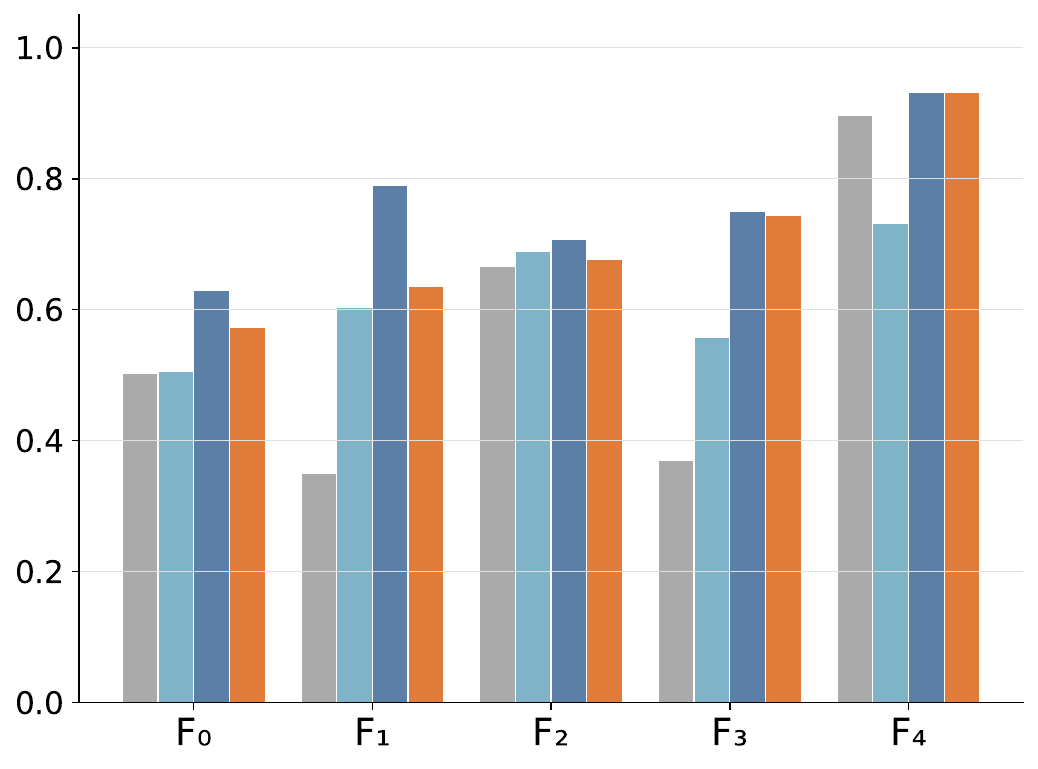}
    \caption{Bertscore\cite{mukherjee2025bleurt}}
    \label{fig:Bertscore}
\end{subfigure}
\hfill
\begin{subfigure}[b]{0.3\textwidth}
    \centering
    \includegraphics[width=\textwidth]{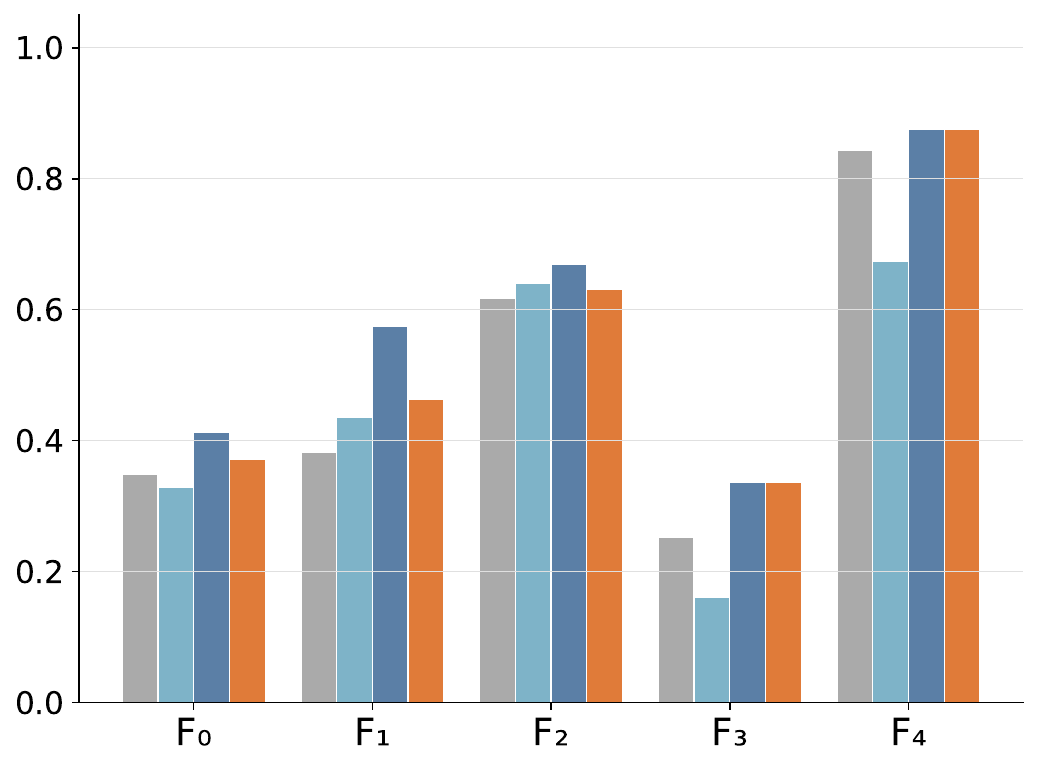}
    \caption{BLEU\cite{papineni2002bleu}}
    \label{fig:bleu}
\end{subfigure}
\hfill
\begin{subfigure}[b]{0.3\textwidth}
    \centering
    \includegraphics[width=\textwidth]{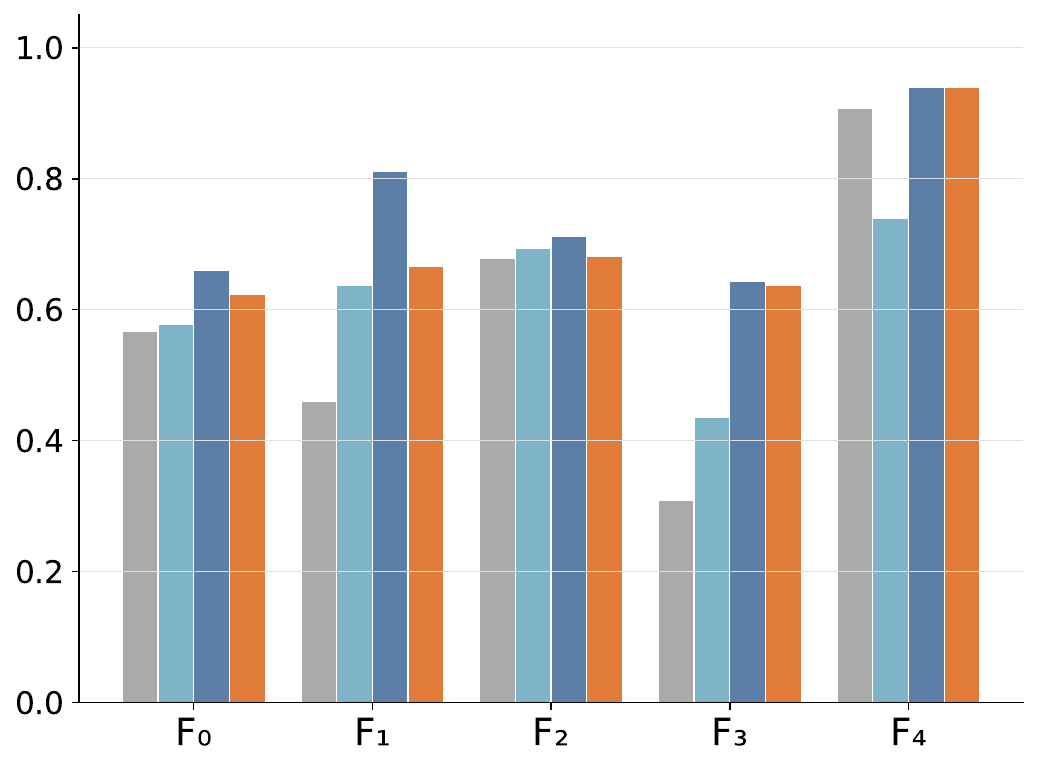}
    \caption{Cosine Similarity\cite{lahitani2016cosine}}
    \label{fig:cosine}
\end{subfigure}
\vspace{0.1cm}
\begin{subfigure}[b]{0.3\textwidth}
    \centering
    \includegraphics[width=\textwidth]{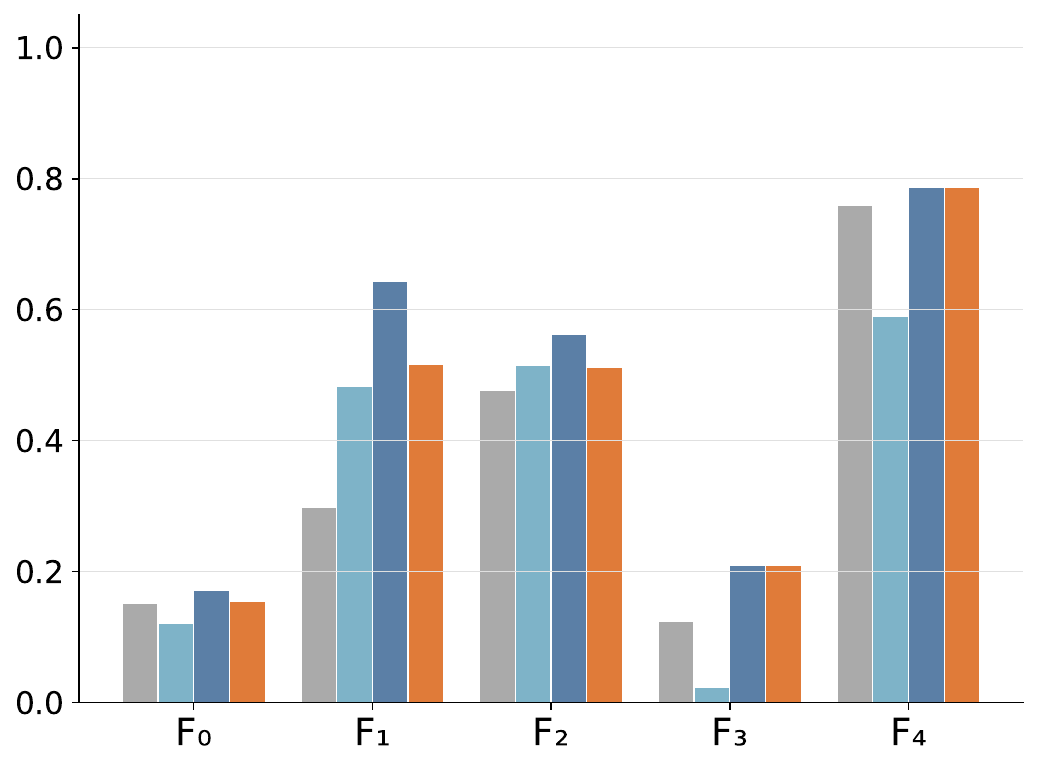}
    \caption{Exact match}
    \label{fig:exact_match}
\end{subfigure}
\hfill
\begin{subfigure}[b]{0.3\textwidth}
    \centering
    \includegraphics[width=\textwidth]{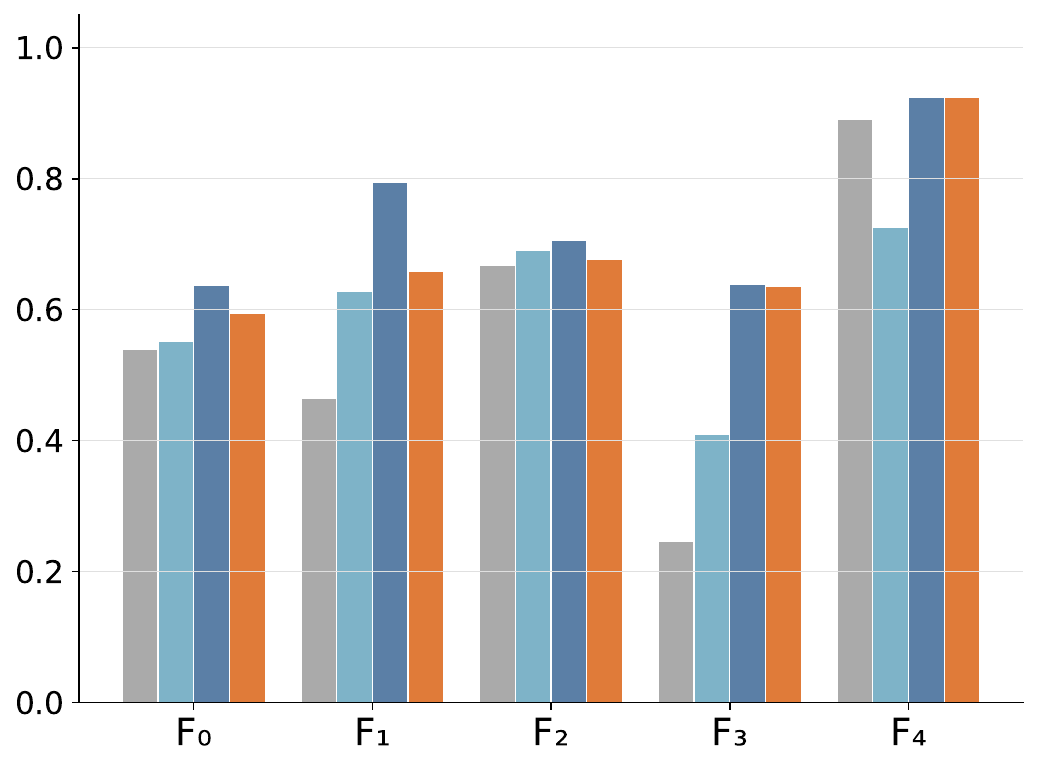}
    \caption{Levenshtein\cite{levenshtein1966binary}}
    \label{fig:Levenshteim}
\end{subfigure}
\hfill
\begin{subfigure}[b]{0.3\textwidth}
    \centering
    \includegraphics[width=\textwidth]{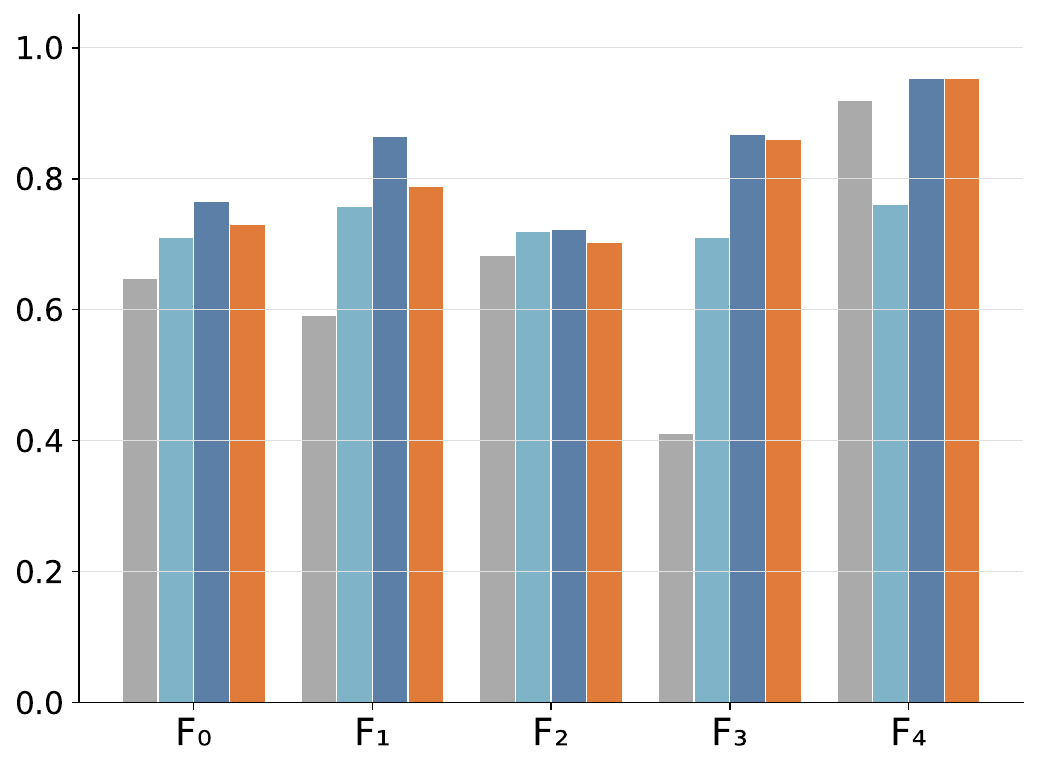}
    \caption{Jaro Winkler\cite{jaro1989advances}}
    \label{fig:jaro_winkle}
\end{subfigure}
\caption{Fidelity of \acro's responses relative to ground truth across six text-based similarity metrics and broken down by $F_i$ impact level; 
\legswatch{purecowrie} Cowrie, \legswatch{ondevicehp} local LLM, \legswatch{cloudhp} Cloud LLM, and \legswatch{hydraorange} \acro.}
\label{fig:fidelity_by_impact}
\end{figure*}

\begin{figure}[t]
\centering
\begin{subfigure}[b]{0.23\textwidth}
    \centering
    \includegraphics[width=\textwidth]{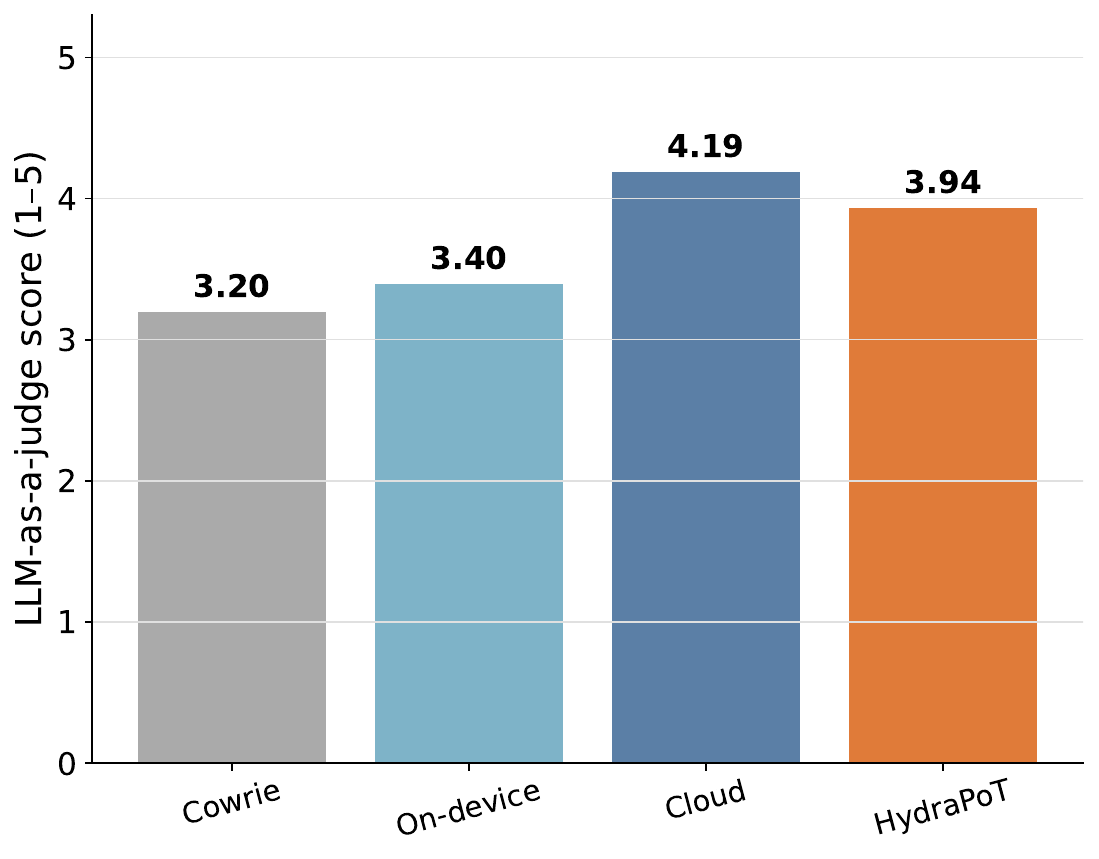}
    \caption{Overall score}
    \label{fig:llm-as-a-judge}
\end{subfigure}
\begin{subfigure}[b]{0.23\textwidth}
    \centering
    \includegraphics[width=\textwidth]{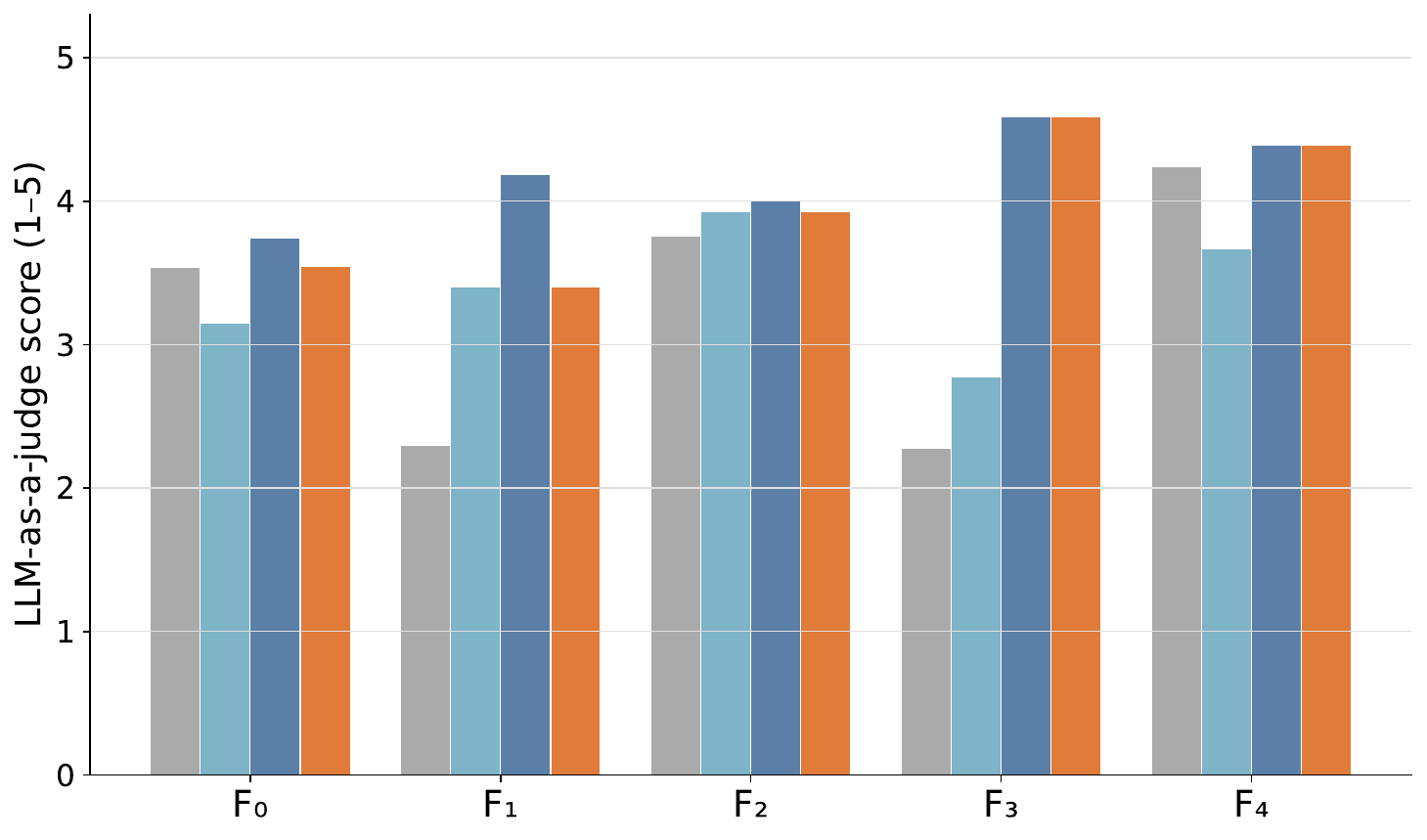}
    \caption{Score per complexity level}
    \label{fig:fi-llm}
\end{subfigure}
\caption{LLM-as-a-Judge fidelity scores. \legswatch{purecowrie} Cowrie-only, \legswatch{pureondevice} local LLM-only, \legswatch{cloudhp} Cloud LLM-only, and \legswatch{hydraorange} \acro under the case-study routing policy.}
\label{fig:llm-fidelity}
\end{figure}

One key benefit of \acro is its configurability that can be used for security practitioners to assign commands of different complexity levels to different honeypot agents according to their desired trade-off between operational cost and deception fidelity. To showcase this benefit, we consider a small-to-medium-sized organization with a moderate (but still limited) budget for accessing cloud-based LLM APIs. 
However, rather than using the cloud LLM for every interaction, the organization reserves it for commands that are more difficult to emulate, while relying on Cowrie or an local LLM for simpler commands.


For this case study, we evaluate using 2,179 commands from 109 attacker sessions in the ML4Net SSH-Shell-Attacks dataset~\cite{botticella2025sshshellattacks}. We configure \acro according to the following routing policy: state-independent $F_0$ commands are handled by Cowrie, state-dependent $F_0$--$F_2$ commands by the local LLM, and $F_3$--$F_4$ by the cloud LLM. 
We evaluate the resulting configuration in terms of response fidelity and operational cost. We compare it with three single-agent baselines that route every command exclusively to Cowrie, the local LLM, or the cloud LLM.

\textbf{Fidelity.}
Following prior work~\cite{malhotra2025llmhoney}, we evaluate response fidelity using six text-based similarity metrics: Cosine Similarity~\cite{lahitani2016cosine}, BERTScore with a pretrained RoBERTa model~\cite{liu2019roberta}, Levenshtein~\cite{levenshtein1966binary}, Jaro--Winkler Similarity~\cite{jaro1989advances}, BLEU~\cite{papineni2002bleu}, and Exact Match. For each command, we use the output produced by executing the same command sequence on an Ubuntu 22.04 virtual machine as the ground truth.

\begin{table}[!t]
\centering
\scriptsize
\caption{Fidelity score metric and results}
\label{tab:fidelity_summary}
\begin{tabular}{@{}lcccc@{}}
\toprule
\textbf{Metric} & \textbf{Cowrie} & \textbf{local} & \textbf{Cloud} & \textbf{\acro} \\
\midrule
Cosine Similarity & 0.574 & 0.594 & \textbf{0.699} & 0.662 \\
BERTScore          & 0.534 & 0.562 & \textbf{0.691} & 0.642 \\
Levenshtein        & 0.548 & 0.572 & \textbf{0.682} & 0.643 \\
Jaro--Winkler       & 0.646 & 0.719 & \textbf{0.795} & 0.766 \\
BLEU                & 0.418 & 0.386 & \textbf{0.487} & 0.450 \\
Exact Match         & 0.254 & 0.226 & \textbf{0.311} & 0.286 \\
\bottomrule
\end{tabular}
\end{table}

Table ~\ref{tab:fidelity_summary}  summarizes the overall fidelity results. Across all six metrics, \acro with the new routing outperforms both Cowrie-only and the local LLM-only \acro, while remaining within 8\% of the cloud LLM-only \acro. 
These results show that \acro retains most of the fidelity provided by the powerful LLM while avoiding the cost of invoking it for every command.

Figure~\ref{fig:fidelity_by_impact} further breaks down the results by impact factor. For each level, \acro achieves fidelity close to that of the agent selected by its routing policy: Cowrie for $F_0$, the local LLM for $F_1$--$F_2$, and the cloud LLM for $F_3$--$F_4$. 

\begin{table}[h]
\centering
\scriptsize
\caption{LLM-as-a-Judge scoring rubric}
\label{tab:judge_rubric}
\begin{tabular}{@{}cp{0.72\linewidth}@{}}
\toprule
\textbf{Score} & \textbf{Meaning} \\
\midrule
5 & Almost the same as a real Linux system. The output is correct, or only differs in small details (such as the hostname, timestamp, or process ID) that could naturally vary on a real machine. \\[4pt]

4 & Looks realistic overall, with only small mistakes in formatting or minor details that most attackers would not notice. \\[4pt]

3 & Mostly correct, but contains noticeable mistakes that an experienced attacker may recognize, such as missing fields or incorrect error messages. \\[4pt]

2 & Partly resembles the expected output, but important information is incorrect or missing, making the response unrealistic. \\[4pt]

1 & Clearly unrealistic. The response is unrelated to the command, contains obvious hallucinations, or is missing output that should normally appear.\\
\bottomrule
\end{tabular}
\end{table}

In addition to the text-based metrics, we evaluate fidelity using an LLM-as-a-Judge~\cite{zheng2023judging}, with GPT-5 ~\cite{openai2025gpt5}. For each command, the judge receives the host profile (hostname, OS, kernel, and architecture), the current \texttt{SYSTEM\_STATE}, the attacker command, and the response generated by the evaluated honeypot, but never the ground-truth output. It scores the response's realism from 1 (\emph{highly unrealistic}) to 5 (\emph{highly realistic}) using the rubric in Table~\ref{tab:judge_rubric}, rather than requiring an exact text match. As a calibration check, the same judge assigns a score of 5 to all ground-truth Ubuntu outputs.

Figure~\ref{fig:llm-as-a-judge} presents the average judge scores. Consistent with the text-based metrics, \acro outperforms the Cowrie-only and lightweight-LLM-only baselines. Its average score is only 0.25 points lower than that of the powerful-LLM-only baseline on the five-point scale, asserting that the proposed routing policy retains most of the realism of the powerful LLM without invoking it for every command. Figure~\ref{fig:fi-llm} further reports the scores by impact factor. For each $F_i$ level, \acro achieves a score close to that of the agent selected by its routing policy, demonstrating that it can combine the fidelity characteristics of heterogeneous agents within one configurable framework.

\textbf{Operational Cost.}
We next evaluate the monetary cost of operating \acro under the case study. For each command, we estimate the marginal cost according to the agent selected by the Honey Router. We treat the computational cost of Cowrie as negligible. 
For the local LLM, we calculate the electricity cost from its measured energy consumption during inference and the local electricity rate;
whereas for the cloud LLM, we calculate the API cost from the number of input and output tokens using the provider's official pricing~\cite{deepseek_pricing}. The total operational cost is then obtained by summing the costs of all commands in the evaluation dataset.

\begin{table}[!t]
\centering
\scriptsize
\caption{Token and API usage}
\label{tab:token_savings}
\begin{tabular}{@{}lrr@{}}
\toprule
\textbf{Configuration} & \textbf{\# Commands Routed to Cloud LLM} & \textbf{Total Tokens} \\
\midrule
Cloud LLM-only baseline & 2,179 & 11,802,012 \\
\acro       & 531   & 3,238,807  \\
\midrule
\textbf{Tokens Saved} & \multicolumn{2}{l}{1,648 fewer cloud-routed commands} \\
 & \multicolumn{2}{l}{8,563,205 (72.6\% reduction)} \\
\bottomrule
\end{tabular}
\end{table}

\begin{table}[!t]
\centering
\scriptsize
\caption{Operational costs}
\label{tab:cost_comparison}
\begin{tabular}{lccc}
\toprule
\textbf{Setup} & \textbf{Cloud Cost} & \textbf{Electricity Cost} & \textbf{Total Cost} \\
\midrule
Baseline: Cowrie-only          & \$0.00 & $\sim$\$0   & \$0.000 \\
Baseline: local LLM-only       & \$0.00 & \$0.017     & \$0.017 \\
Baseline: Cloud LLM-only       & \$1.98 & $\sim$\$0         & \$1.98 \\
\acro         & \$0.54 & \$0.0015     & \$0.5415 \\
\bottomrule
\end{tabular}
\end{table}

As shown in Tables~\ref{tab:token_savings} and~\ref{tab:cost_comparison}, the routing policy reduces cloud-processed tokens by 72.6\% and operational cost by 3.7$\times$, from \$1.98 for the cloud-LLM-only baseline to \$0.54 for \acro. Extrapolated to realistic workload of 1,000 commands per day, this is about \$91 per year instead of \$332, resulting in roughly \$241 in annual monetary savings. Combined with the fidelity results in Table~\ref{tab:fidelity_summary}, \acro thus achieves fidelity close to the powerful-LLM-only honeypot at about a quarter of its cost, which is particularly appealing to small-to-medium-sized organizations with limited budgets.

\section{Other Features of \acro}

Besides enabling a configurable cost--fidelity trade-off, \acro provides two additional features: plugin-based support for user-defined command simulation and seamless integration with SIEM platforms.
Together, these features extend \acro beyond a standalone honeypot into a ready-to-use threat-intelligence platform.

\textbf{User-Defined Plugin-Based Command Simulation.}
\acro provides a plugin-based interface that allows security practitioners to define custom command handlers and environment-specific responses for commands that can be simulated deterministically. These plugins can be tailored for specific software, services, files, and system behavior of a target production environment, thereby improving the realism and relevance of collected attacker interactions.

Security practitioners may also customize the Honey Router to determine which commands are handled by these plugins and which are forwarded to the lightweight or powerful LLM agents. 
This provides fine-grained control over command processing while allowing organizations to adapt \acro to their deployment environment and investigation objectives.


\textbf{Security Information and Event Management Integration (SIEM).}
\label{subsec:siem}
\acro provides a SIEM-style dashboard, designed to support Security Operations Center (SOC) teams. 
Using data collected by \acro's State/Log Manager, the dashboard displays attacker interactions through a live terminal feed. 
It also summarizes relevant information, including observed attack techniques, command impact levels based on the $F_i$ score, source IP addresses, attack activity over time, anomalous requests, session-level activity, the distribution of commands across agents, and the operational cost of cloud-based LLM usage. 
This integration allows security practitioners to monitor ongoing sessions and use the collected interactions for subsequent threat-intelligence analysis.


\section{Conclusion}
In this paper, we presented \acro, a configurable multi-agent honeypot architecture that combines a traditional honeypot, a local LLM, and a cloud LLM under a unified honeypot framework. Unlike existing AI-powered honeypots that rely on a single LLM agent, \acro allows different agents to collaborate within the same attacker session while maintaining a shared system state and interaction history. 
We envision that \acro can be useful for organizations to enable customized routing policies based on their own requirements, providing a practical trade-off between interaction fidelity, operational cost, and resource usage.

Future work will include human evaluation and real-world deployment to assess its fidelity, reliability, and performance under realistic attack traffic. Another interesting direction is mitigation to prompt injection, which remains unexplored in the context of cost-aware LLM-assisted honeypot systems.

\textbf{Acknowledgments.} This research is funded by National Research of Thailand (NRCT) and National Science and Technology Development Agency (NSTDA).
It is also partially supported by College of Computing, Prince of Songkla University, Phuket Campus. 
Finally, we thank the anonymous AINTEC reviewers for their helpful feedback.




\bibliographystyle{ACM-Reference-Format}
\bibliography{references}

\end{document}